\documentclass[12pt]{article}
\usepackage{epsf,amsmath,amssymb}

\newcommand{\bea}{\begin{eqnarray}}

\def\be#1\ee{\begin{equation}#1\end{equation}}
\def\up{\uparrow}
\newcommand{\om}{\omega}

\newcommand{\Z}{\mathbb{Z}}
\newcommand{\bR}{\mathbb{R}}

\renewcommand{\H}{\mathbb{H}}
\newcommand{\1}{\mbox{1\hspace{-.4ex}I}}

\newcommand{\cB}{{\cal B}}  
\newcommand{\cC}{{\cal C}}

\newcommand{\cH}{{\cal H}}

\newcommand{\cM}{{\cal M}}  

\newcommand{\cP}{{\cal P}}

\newcommand{\cV}{{\cal V}}

\def\tr{{\rm tr}}

\def\la{\langle}
\def\ra{\rangle}

\newcommand{\CMP}[1]{Comm.\ Math.\ Phys.\ {\bf #1}}
\newcommand{\PR}[1]{Phys.\ Rev.\ {\bf #1}}
\newcommand{\PRL}[1]{Phys.\ Rev.\ Lett.\ {\bf #1}}

\begin{document} 
\begin{flushright} 
MPP-2010-147\\
\end{flushright} 
\vglue-13cm
\begin{center} 
{\Large The Strange World of Non-amenable Symmetries} 
\footnote{Talk given  at the International Conference ``Mathematical 
Quantum Field Theory\\ 
and Renormalization Theory'', Nov.~26 to Nov.~29, 2009, Kyushu 
University, 
Fukuoka, Japan}
\vglue5mm 
Erhard Seiler
\end{center} 
\begin{center}
{\it Max-Planck-Institut f\"ur Physik\\  
(Werner-Heisenberg-Institut)\\ 
F\"ohringer Ring 6, 80805\\ 
Munich, Germany\\  
e-mail: ehs@mppmu.mpg.de} 
\end{center}
\bigskip 
\nopagebreak 

\begin{abstract} 
\noindent 
Nonlinear sigma models with non-compact target space and non-amen-able 
symmetry group were introduced long ago in the study of disordered 
electron systems. They also occur in dimensionally reduced quantum 
gravity; recently they have been considered in the context of the AdS/CFT 
correspondence. These models show spontaneous symmetry breaking in {\it any} 
dimension, even one and two (superficially in contradiction with the 
Mermin-Wagner theorem) as a consequence of the non-amenability of their 
symmetry group. The low-dimensional models show other peculiarities: 
invariant observables remain dependent on boundary conditions in the 
thermodynamic limit and the Osterwalder-Schrader reconstruction yields a 
non-separable Hilbert space. The ground state space, however, under quite 
general conditions, carries a unique unitary and continuous 
representation. The existence of a continuum limit in 2D is an open 
question: while the perturbative Renormalization Group suggests 
triviality, other arguments hint at the existence of a conformally 
invariant continuum limit at least for suitable observables.

This talk gives an overview of the work done during the last several
years in collaboration first of all with Max Niedermaier, some of it also 
with Peter Weisz and Tony Duncan \cite{NieSei1,DNS,NieSei3,NieSeiW,NieSei4}. 
\end{abstract}
\vskip2mm
\section{Introduction}
\subsection{What are non-amenable symmetries?}

The concept of amenable groups was introduced by J. von Neumann in 1929; 
it can be described as follows: let $ {\cC(G)}$ be the space of continuous 
bounded functions on $G$; then a {\it mean} $m$ is a positive (hence 
continuous) linear functional on $\cC(G)$ (or on $ {L^\infty(G)}$) with 
$ {m(\1)=1}$. Put differently: $m$ is a {\it state} on the commutative 
$C^*$ algebra $\cC(G)$ (or ${L^\infty(G)}$). Since the group $G$ has a 
natural left action on functions, it makes sense to speak about invariance 
of such a mean.

{\it Definition:} The group $G$ is non-amenable if there is no mean on  
${\cC(G)}$ which is invariant under $ G$.

A well-known fact is that noncompact semisimple Lie groups are 
nonamenable \cite{paterson}.

The concept can be generalized to homogeneous spaces $G/H$ by using the 
the algebra $\cC(G/H)$ instead of $\cC(G)$. One also speaks of 
(non-)amenable actions of a group $G$ on a general $G$-space $\cM$: the 
left action of $G$ induces an action on $ \cC(\cM)$ and non-amenability 
means nonexistence of a mean invariant under $G$ on $\cC(\cM)$.

Bekka \cite{bekka} has extended the definition to that of amenable unitary 
representations $\pi$ on a Hilbert space $\cH$ as follows: $\pi$ is called 
amenable if there is a state on $\cB(\cH)$ which is invariant under $\pi$. 
In this context the following result is important: if $G$ is simple, 
noncompact, connected, with finite center, rank $ {>1}$, the trivial 
representation is the only amenable one.

\subsection {Physics motivations} 

(1)
Nonlinear $\sigma$ models with hyperbolic target space -- the prototype of 
a non-amenable symmetric space -- were introduced 1979 by Wegner 
\cite{wegner} to describe the conductor-insulator transition in disordered 
electron systems. Since then there has been a lot of activity, see for 
instance \cite{WS,HJKP,Hik}. Later Efetov \cite{efetov83, efetov97} and 
Zirnbauer \cite{zirn} introduced the supersymmetric version of that model 
as a better description of the electron system. This line of research was 
continued more recently in \cite{SZ, DSZ, DS}. 

(2) 
Some `warped' versions of nonlinear $\sigma$ models with hyperbolic 
target space arise in dimensionally reduced gravity and its quantization 
\cite{breilo,hollmann,NR}. 

(3)
Not surprisingly, these models also appear in the context of string 
theory; string theorists think of hyperbolic space as `Euclidean Anti-de 
Sitter space' \cite{PST,FG}.

\section{Quantum Mechanics on hyperbolic spaces}

Insight into the peculiarites of non-amenable symmetries is easiest to 
obtain by studying quantum mechanics on hyperbolic spaces. Hyperbolic 
space can be described as a hyperboloid imbedded in Minkowski space with 
the metric induced by the ambient space:
\be
\H_N\equiv SO_o(1,N)/SO_o(N)\equiv G/K=\{n\in \bR^{N+1}|
n\cdot n=1,\ n_o>0\}\,,
\ee
where $n\cdot n'\ = \ n_o n'_o-\vec n \cdot \vec n'$. 

\subsection{One particle}

Let $\Delta$ denote the Laplace-Beltrami operator on $\H_N$. The free 
one particle Hamiltonian, acting on $L^2(\H_N,d\Omega)$ where $d\Omega$ 
is an invariant measure on $\H_N$, is then
\be
H=-\Delta\ge 0\,.
\ee
This Hamiltonian is diagonalized using the Mehler-Fock transformation 
\cite{dym}; it reveals that the spectrum of $H$ is absolutely continuous, 
covering the interval $[(N-1)^2/4,~\infty)$; there is no spectrum 
in the interval $[0,(N-1)^2/4)$, even though there are bounded 
eigenfunctions for every value in that interval (the supplementary 
series). Introducing a spectral parameter $\omega$ running from 0 to 
$\infty$, we have the spectral resolution
\be
L^2(\H_N,d\Omega)=\int_0^\infty dP(\omega) \cH_\omega;\quad 
H\psi= \int_0^\infty dP(\omega) \left(\frac{1}{4}(N-1)^2+\om^2)\psi
\right)\,.
\ee
Only the principal series appears; the spectrum is infinitely degenerate 
because all representations in that series are infinite dimensional. Of 
course there is no normalizable ground state vector; instead we have a 
`ground state space' corresponding to $\om=0$ and spanned by `generalized 
ground states' (functions in $\cC(\cM)$ but $\notin L^2$) of the form
\be
\cP_{-1/2}^{1-N/2}(gn\cdot n^\up),\quad g\in SO_o(1,N)\,,
\ee
and linear combinations thereof, where $\cP_{-1/2}^{1-N/2}$ are Legendre 
functions. 

A different scalar product, produced by the Osterwalder-Schrader (OS) 
reconstruction makes these ground states normalizable. They then generate 
a Hilbert space of ground states carrying a special unitary irreducible 
representation $\sigma_0$. But the main point is this:

\begin{center}
{\bf There is no invariant ground state\\ 
Spontaneous symmetry breaking (SSB) takes place!}
\end{center}

\subsection{$\nu$ particles: separation of `center of mass'}

When we consider a $\nu$ particle system interacting via translation 
invariant potentials in Euclidean space, the first step is always to 
separate out the free center of mass motion. Here we consider $\nu$ 
particles on $\H_N$ with a potential invariant under the symmetry group 
$G=SO_0(1,N)$, and again we would like to to find a way to extract the 
rigid motions. This requires some tricks.

Our Hilbert space is now $\cH=L^2(\cM)$ ($\cM=\H_N^\nu$) and the 
Hamiltionian is
\be
H=-\sum_{i=1}^n \Delta_i+ \sum_{i<j}V(n_i\cdot n_j)\equiv H_0+\cV\,.
\ee
Let $\ell_{\cM}$ be the unitary representation of $G$ on $\cH$
induced by the left diagonal action of $G$, representing rigid motions of 
the particle system. Clearly 
\be
[H,\ell_\cM(G)]=0\,,
\ee
We now turn the left diagonal action on $\cM$ into a right action on a 
different manifold $\cM_r$ ins such a way that only one `center of 
mass' variable is affected. First we define
\be
\tilde\cM_r\equiv G\times \cH^{\nu-1}
\ee
and an injective but not surjective map  $\tilde \phi:\cM\to\tilde\cM_r$ 
given by
\be
\tilde\phi(n_1,\ldots, n_n)=(g_s(n_1)^{-1},g_s(n_1)^{-1}n_2,\ldots
g_s(n_1)^{-1}n_\nu)\,,
\ee
where $g_s$ is a function (global section) $\H_N\to G$ such that
$n=g_s(n^\up)$. $g_s$ is obviously only determined up to $g_s\to 
g_sk^{-1}, k\in K$. Let $d_\ell(K)$ be the left diagonal action of $K$ on 
$\tilde\cM_r$ and define
\be
\cM_r=\tilde\cM_r/d_\ell(K)\,.  
\ee
$\tilde \phi$ projects to a well-defined map $\phi:\cM\to\cM_r$ and this 
$\phi$ does the job of converting the left diagonal action $d_\ell(G)$ on 
$\cM$ into a right action $r(G)$ on $\cM_r$ acting only on the first 
entry:
\be
r(g')[(g,n_1,\ldots,n_\nu)]=[gg',n_1,\ldots,n_\nu)]\,,
\ee
\be
\phi\circ d_\ell = r\circ\phi\,. 
\ee
$\phi$ induces a unitary map $\Phi$ between the corresponding Hilbert 
spaces $L^2(\cM_r)$ and $L^2(\cM)$; the latter can be viewed as the 
subspace of $L^2(\tilde\cM_r)$ invariant under the unitary map induced
by $d_\ell(K)$. 

The right action of $G$ on the first entry of $\cM_r$ induces a unitary 
representation $\rho(G)$ of the rigid motions:
\be
\rho=\Phi^{-1}\circ\ell_\cM\circ\Phi\,
\ee
and $\rho(G)$ commutes with $\ell_r(K)$.

\subsection{The ground state representation}

The harmonic analysis of $\rho$ is the analogue of ~the decomposition
according to the center of mass momentum in flat space. The Hilbert space
$\cH=L^2(\cM_r)$ decomposes into a direct integral of irreps 
\be
\cH=\int_{\widehat G_r}^\oplus d\nu(\sigma) \cH(\sigma)\,,
\ee
where $\widehat G_r$ is the restricted dual of $G$, which is the union of 
the principal and the discrete series (see \cite{folland})
\be
\widehat G_r=\widehat G_p\cup \widehat G_d\,;
\ee
$d\nu$ arises from the Plancherel measure. 

The Hamiltonian on $\cH$ is $H_r=\Phi^{-1}\circ H\circ \Phi$; we drop the 
subscript $r$ from now on. Because $H$ commutes with $\rho$, it can also 
be resolved into fiber Hamiltonians 
\be
H=\int_{\widehat G_r}^\oplus d\nu(\sigma) h(\sigma)\,,
\ee
which is analogous to the resolution of a $\nu$ particle Hamiltonian 
according to the c.m. momentum in flat space.

We conjecture that generally $d\nu$ is carried by $\widehat G_p$ alone,
but we can prove only that 
\be
\inf_\sigma \inf\ {\rm spec}\ h(\sigma) \notin {\widehat G_d}\,.
\ee
This is seen most easily under a certain compactness condition on the 
interaction $\cV$, namely
\be
\tr\ e^{-t(H_o+\cV+\cV_1)}\le \tr\ (e^{-tH_o}e^{-t(\cV+\cV_1)})<\infty\,. 
\ee

This implies that the fiber Hamiltonians $h(.)$ have discrete spectrum;  
for $\sigma\in\widehat G_d$ the ground state of the fiber Hamiltonian 
would give rise to a (normalizable) eigenfunction of $H$; because 
$\sigma$ is not the trivial representation, this ground state could not 
be unique. This leads to a contradiction with the Perron-Frobenius 
theorem.  We can show furthermore that the ground state representation is 
always $\sigma_0$, the special representation found for the one particle 
case. Details can be found in \cite{NieSei3}, where, however, we deal with 
a discrete time evolution given by a transfer matrix.

The ground state representation $\sigma_0$ is universal, and the fact that 
it is nontrivial means again that there is SSB.
\section{Statistical mechanics / lattice quantum field theory}
\subsection{Action, Gibbs state}

We consider configurations of `spins' given by mapping each site $x\in 
\Lambda\subset \Z^d$ to a $n(x)\in\H_N$. The Gibbs measure is formally 
given by
\be
\exp(-\beta S)\prod_x d\Omega(x)\,,
\ee
with (for instance)
\be
S=\sum_{\la xy\ra} n(x)\cdot n(y)\,.
\ee
To make the Gibbs measure normalizable, `gauge fixing' is needed. The 
simplest choice is to fix a spin at the boundary of the finite lattice 
$\Lambda$.  

\subsection{Spontaneous symmetry breaking}

If in the thermodynamic limit $\Lambda\nearrow\Z^d$ the Gibbs state is not 
invariant under $G$, we speak of SSB. Non-amenability {\it enforces} SSB,
because if there were a symmetric Gibbs measure, it would automatically 
induce an invariant mean on the functions of a single spin. This holds 
independent of the dimension $d$ or any other details (type of lattice, 
action).

\begin{center}
{\bf SSB is unavoidable!}
\end{center}

Note that the Mermin-Wagner theorem is not in conflict with this finding: 
it forbids SSB only for {\it compact} symmetry groups in dimensions 1 and 
2.

\subsection{The hyperbolic spin chain}

For $d=1$ the problem can be solved analytically to a large extent 
\cite{NieSei1}. `Gauge fixing' is done by fixing the spin at the left hand 
end of the chain, say
\be
n(-L)=n^\up\,.
\ee
As the general considerations require, SSB occurs in the form that the system 
remembers the orientation  of the spin $n(-L)$ even in the limit 
$L\to\infty$. A concrete `order parameter' that shows this is
\be
T_e(n(0)):=\tanh(n(0)\cdot e)\,,
\ee 
where $e\cdot e=-1$. In \cite{NieSei1} it is shown that 
\be
\lim_{L\to\infty} \la T_e(n(0)) \ra_{bc}=1-\frac{2}{\pi}
\cos^{-1}\left(\frac{e\cdot n^\uparrow}
{\sqrt{1+(e\cdot n^\uparrow)^2}}\right)\,.
\ee
Two facts are might be unexpected:\\
(1) 
even the expectation values of invariant functions, such as $n(0)\cdot 
n(x)$ remain dependent on b.c. in the thermodynamic limit.

(2) The OS reconstruction of a Hilbert space from the correlation 
functions yields a {\it non-separable} space and discontinuous 
representations, except for the ground state space. This is related to the 
fact that the construction always produces a {\it normalizable} ground 
state, even though there is none in the original $L^2$ space, so in some 
sense the OS reconstruction renormalizes the scalar product.

\subsection{Two or more dimensions}

Not very much is known rigorously beyond the fact of SSB.  The models do 
not have high temperature expansions; presumably they are massless at all 
temperatures.

Spencer and Zirnbauer \cite{SZ} have shown, however, that in dimension 3 
or more at low temperature there is a `stronger version' of SSB that 
presumably is not true in dimensions 1 or 2 or at high temperatures; 
namely the quadratic fluctuations away from the mean `magnetiztion' 
have a finite expectation value.

In \cite{DNS} we carried out some detailed numerical simulation of the 
model in $d=2$ with a different (translation invariant) gauge fixing. We 
found

(1) The explicit symmetry violations due to the gauge fixing disappear in 
the thermodynamic limit; this is seen by verifying Ward identities

(2) SSB is seen by looking at $\la T(e)\ra$  as above.

(3) The thermodynamic limit for invariant observables seems to exist.

\section{Quantum field theoretic considerations}

\subsection{Peculiarities of the Osterwalder-Schrader reconstruction}

As in the hyperbolic spin chain, the OS reconstruction will presumably 
always lead to a nonseparable Hilbert space. This is a consequence of the 
fact that the construction always yields a normalized ground state, even 
though the spectrum of the transfer matrix is most likely continuous.

This somewhat unphysical feature might be avoided by restricting the space 
of observables. For instance one might restrict attention to only a 
certain component of the spins, or maybe a special (horospherical) 
coordinate and functions of it. In this way one would of course lose the 
signal of SSB.

\subsection{Existence of a continuum limit?}

Consideration of the perturbative one loop Renormalization Group 
\cite{friedan, lott} yields essentially the Ricci flow, indicating that 
the model is {\it infrared asymptotically free}. In the case at hand, 
however, this is counterintuitive: if the long-distance fluctuations 
become Gaussian, as infrared asymptototic freedom would predict, this 
would mean that they doen't feel the curvature of the target manifond. But 
in the infrared the fluctuations necessarily cover the target space over 
large distances and therefore should become extremely sensitive to the 
curvature.

But if the conventional wisdom is right, it would suggest that there is no 
nontrivial continuum limit of 2D nonlinear $\sigma$ models whose target 
space has negative curvature; the situation would be similar to the 
QED$_4$ or $\phi^4_4$ quantum field theories, which are believed to be 
trivial (i.e. Gaussian) in the continuum limit.

A counterpoint has been provided long ago by Haba \cite{haba}, who by a 
formal calculation of the 2D hyperbolic $\sigma$ model concluded that it 
corresponded to a conformal quantum field theory with central Virasoro 
charge $c=1$, as long as $\beta> 1/3\pi$ (it should be noted, however, 
that he considered only correlations of a so-called horospherical 
coordinate on the hyperbolic plane). It would be very interesting to know 
if this formal calculation can be justified.

\subsection{Axiomatic considerations}

When considering possible continuum limits, it is worthwhile to pause and 
think how such limits could possibly look, in agreement with the axiomatic 
structure of quantum field theory.

One thing becomes clear immediately: it is not possible to have a 
multiplet of quantum fields $\phi_i$ transforming under the non-unitary 
vector representation of $G=SO(1,N)$ with unbroken symmetry (as one would 
expect naively if the $\phi_i$ are continuum fields arising from 
renormalizing the basic spin components $n_i$).

The reasoning goes like this: an unbroken symmetry means that there is a 
unitary representation $U(.)$ of the symmetry group $G$ leaving the 
vacuum state invariant and transforming the fields according to the vector 
representation, i.e.
\be
U(g)^{-1}\phi_i(x)U(g)=\sum_j\left(g^{-1}\right)_{ij}\phi_j(x)\,. 
\ee
This leads to a conflict with the positive metric in Hilbert space when 
considering orbits of $U(.)\psi$: let $\phi_0(f),\phi_1(f)$ be field 
components smeared with a test function $f$. Then by the unitarity of 
$U(.)$
\be
\Vert\left(\phi_o(f)\  {\rm ch}\  t +\phi_1(f)\  {\rm sh}\  t\ 
\Omega\right)\Vert^2=
\Vert \phi_o(f)\Omega\Vert^2\quad \forall t\,,
\ee
which  is impossible unless all $\phi_i=0$.

Possible alternatives are: 

(1) There is SSB, hence no unitary represntation $U(.)$ of $G$ (see for 
instance{blot}\\
(2) There is an infinite multiplet of fields, transforming according to a 
unitary representation of $G$ -- this could, however, not correspond to a 
continuum limit of the lattice model\\
(3) A Quantum Field Theory arises only for a subset of fields. The 
symmetry is then not visible. Haba's computation suggests that this might 
be the right scenario.

\section{Conclusions, open questions}

(1) We have found a certain universal ground state representation both in 
Quantum Mechanic and Lattice field theory (in a finite spatial volume).

(2) There is always SSB; the Mermin-Wagner theorem does not apply.
 
(3) In a potential continuum limit also Coleman's version of the  
Mermin-Wagner theorem would not apply because the currents needed for this 
argument don't have thermodynamic and continuum limits.

(4) In $D\ge 3$ for large $\beta$  there is SSB of the conventional kind: 
with large fluctuations suppressed\cite{SZ}.

(5) There is probably no mass gap, but a proof is lacking.

(6) In 2D infrared asymptotic freedom is suggested by perturbation theory, 
but there is no proof; the existence of a continuum limit remains an 
unsolved question.


\end{document}